\documentclass{elsart}
\input psfig

\begin{document}
\runauthor{Coppi and Aharonian}
\begin{frontmatter}
\title{Understanding the Spectra of TeV Blazars: Implications for
the Cosmic Infrared Background}
\author[Paestum]{Paolo S. Coppi\thanksref{NASA}}
\author[Rome]{Felix A. Aharonian}
\address[Paestum]{
Department of Astronomy, Yale University, P.O.
Box 208101, New Haven, CT 06520-8101, USA}
\address[Rome]{Max-Planck-Institut f\"ur Kernphysik, Postfach 103980, D-69029 
Heidelberg, Germany}
%\centerline{To appear in Astroparticle Physics (Proceedings of the
%VERITAS Workshop)}
\thanks[NASA]{Partially supported by NASA grant NAG 5-3686. To appear
in Astroparticle Physics (Proceedings of the VERITAS Workshop on the
TeV Astrophysics of Extragalactic Objects).}

\begin{abstract}
\def\mathnew{\mathsurround=0pt}
\def\simov#1#2{\lower .5pt\vbox{\baselineskip0pt \lineskip-.5pt
        \ialign{$\mathnew#1\hfil##\hfil$\crcr#2\crcr\sim\crcr}}}
\def\simgreat{\mathrel{\mathpalette\simov >}}
\def\apgt{\simgreat}
\def\simless{\mathrel{\mathpalette\simov <}}
\def\aplt{\simless}
\def\g{{$\gamma$}}

With the arrival of powerful, ground-based \g-ray detectors,
we can now begin to seriously probe, 
via simultaneous X-ray/TeV observations,
the origin of the \g-ray emission in the blazars Mrk 421 and 501.
If the synchrotron-Compton emission model turns out to
work, then we know that the same electrons are responsible 
for both the X-ray and the \g-ray emission of these objects.
In this case, we show that we can use their observed X-ray spectra
to robustly estimate their intrinsic \g-ray spectra. Among blazars, 
Mrk 421/501 are particularly well-suited for this task
because the Compton scattering
which produces their TeV \g-rays is likely to be in the Klein-Nishina
limit, where the outgoing photon has an energy insensitive
to the incoming photon energy. With a better handle on their
intrinsic TeV spectra, we can then begin to search
for evidence of absorption due to \g-ray pair production
on diffuse infrared background radiation. We discuss some of the pitfalls
that arise when one attempts to do this without knowing the intrinsic
spectrum. Even though Mrk 421/501 are very nearby, the emission
of these sources extends to sufficiently high
energies ($\simgreat 20$ TeV in Mrk 501) that we may
nevertheless be able to derive interesting constraints
on the infrared background. If correct, the combination of the
COBE 140 $\mu$ detection and the  measurement of Mrk 501's
spectrum out to beyond $\sim 20$ TeV  rules out
conventional galaxy evolution and star formation scenarios,
implying that much of the star formation in the Universe 
indeed occurs at early times in highly obscured sources
that have been missed until now.

\end{abstract}
\end{frontmatter}

\def\mathnew{\mathsurround=0pt}
\def\simov#1#2{\lower .5pt\vbox{\baselineskip0pt \lineskip-.5pt
        \ialign{$\mathnew#1\hfil##\hfil$\crcr#2\crcr\sim\crcr}}}
\def\simgreat{\mathrel{\mathpalette\simov >}}
\def\apgt{\simgreat}
\def\simless{\mathrel{\mathpalette\simov <}}
\def\aplt{\simless}
\def\g{{$\gamma$}}

%\author{Paolo S. Coppi}
%\affil{}
%\authoremail{coppi@astro.yale.edu}
%\author{Felix A. Aharonian}
%\affil{ }
%\authoremail{aharon@fel.mpi-hd.mpg.de}

\section{Introduction}

The last years have seen a revolution in ground-based
\g-ray detectors. We can now detect the
spectra of nearby TeV blazars like Mrk 421 and
501 out to $\sim 20$ TeV, and 
%%no. 1
during the strongest  flares  (with observed TeV fluxes up to 10 times that
of the Crab),  we can follow fluctuations
in these spectra on  timescales down to the shortest ones likely in these 
objects. 
This represents a unique
opportunity.  Using these detectors
in combination with X-ray satellites like ASCA, SAX, and RXTE,
we can now begin to simultaneously follow all significant
X-ray/\g-ray variations in a blazar's emission (e.g., see 
the contribution by Takahashi in
this proceedings). This will provide the most stringent test 
yet for the synchrotron-Compton (SC) blazar emission model
(see, e.g.,  \cite{sikora} and \cite{cracow} for reviews of current
emission models and controversies). In this paper,
we will argue that such a test is crucial in helping  us
reach one of the ``holy grails''
of TeV astronomy: the detection of absorption in
blazar \g-ray spectra due to \g-ray pair production on the 
low energy diffuse extragalactic background radiation (DEBRA).
As discussed in detail by several authors in this proceedings 
(see the contributions of Primack, Stecker, and Biller
and references therein), a
 strong constraint on the amount of 
absorption is very exciting because it constrains the
density of the target infrared/optical DEBRA photons responsible
for it.  The DEBRA at these energies is most probably
redshifted stellar and dust emission and thus
contains important information on galaxy evolution and cosmology.
It is hard to measure
by other means, especially in the $\sim 5-50\mu$ range,
because of the large Galactic and solar system
foregrounds present. In \S 2 below, we briefly review what
sort of absorption effects one should expect to see in 
objects like Mrk 421 and Mrk 501. We then show that it is difficult
to constrain  absorption at $\sim 1-10$ TeV without
knowing in detail the  shape
of the intrinsic, unabsorbed \g-ray spectrum. (Note that since blazars
are so variable, one really needs to know the
shape of the {\it instantaneous} \g-ray spectrum.) In particular,
even though Mrk 421/501 are nearby,
their spectra may already be significantly absorbed
(by a factor up to 2!) at 3 TeV.  Therefore, the 
lack of a sharp cutoff in the spectra of both Mrk 421 \cite{Zw97}
and Mrk 501  \cite{Ah9799,Sam98,TA,CAT} up to 10 TeV does not allow us to
unambiguously extract information on DEBRA, although these data
probably are enough to rule out some of the
more exotic DEBRA models \cite{Biller}.   
We then note that the extension of the 
HEGRA measurement of the Mrk 501 spectrum to at least 
$\sim 20$ TeV (Konopelko, this proceedings) may be extremely 
important. Making rather minimal assumptions about
the intrinsic \g-ray spectrum at these energes, we
obtain a strong constraint on the DEBRA 
intensity at $\sim 10-60 \mu.$ All recently
published DEBRA models either run into trouble with this constraint
or significantly underpredict the flux detected by COBE at 
$140 \mu.$  If the COBE detection is correct,  this
implies the sources responsible for the far IR DEBRA are 
qualitatively different from the typical ones we see today.
Most of the star formation in the early Universe must occur 
in highly obscured,  dusty environments.
In \S 3, we show
that if the SC model works during a large flare 
(where the emission from a single region may dominate), 
then we can  use the observed  X-ray spectrum  to 
robustly predict the intrinsic TeV spectrum. Then, and
only then, we can try to look for absorption below $\sim$ 10
TeV.

\section{ Gamma-Ray Pair Production on Diffuse Background Radiation }

To obtain
the mean free path for a \g-ray of energy $E_\gamma,$ 
one must in general convolve the DEBRA photon number distribution,
$n(\epsilon),$  with the pair production cross-section.
However, this cross-section is 
peaked, and for nearby ($z\ll1$) HBLs
and almost all plausible DEBRA shapes,  
over half the interactions occur on
DEBRA target photons
with energies $\epsilon = 0.5-1.5 \epsilon_\ast,$ where
$\epsilon_\ast =
4m_e^2c^4/E_\gamma\approx 1.04 (E_\gamma/1{\rm TeV})^{-1}{\rm\  eV}.$ 
To accuracy better than $\sim 40\%,$
we can thus approximate
the absorption optical depth as
$$\tau_{\gamma\gamma}(E_\gamma) \approx 0.24 ({E_\gamma
\over \rm 1 TeV})({u(\epsilon_\ast) \over
10^{-3} \rm eV cm^{-3}})({z_s \over 0.1}) h_{60}^{-1}.$$
Here $u(\epsilon_\ast)=\epsilon_\ast^2
n(\epsilon_\ast)$ is the typical energy density in a
energy band centered on $\epsilon_\ast,$  $h_{60}$ is
the Hubble constant in units of $60 {\rm km}{\rm s}^{-1}
{\rm Mpc}^{-1},$ and $z_s$ is the source redshift.
If $I_0 (E_\gamma)$ is
the intrinsic source spectrum,  the corresponding observed spectrum
is then $I(E_\gamma)=I_0(E_\gamma)\exp(-\tau_{\gamma\gamma}).$ 
Note that if the DEBRA spectrum near $\epsilon_{\ast}$
can be approximated by a power law, $n(\epsilon) 
\propto \epsilon^{-\alpha_\ast},$ then $\tau_{\gamma\gamma}$
goes as $E_\gamma^{\alpha_{\ast}-1}.$
Connecting
the COBE far IR  measurements to the latest UV background
estimates, one gets a crude DEBRA spectral index $\alpha \sim 2$ 
(e.g., see \cite{dwek} for a good compilation of the latest DEBRA
 observations and models). To zeroth order, then, $\tau_{\gamma\gamma} 
\propto
E_\gamma,$ and the observed spectrum
should  be $\sim I_0(E_\gamma)\exp(-E_\gamma/E_c)$ where 
the cutoff energy $E_c$ is set by
$\tau_{\gamma\gamma}(E_c)=1.$ Interestingly, this is exactly
the type of shape seen in HEGRA observations of
Mrk 501 (Konopelko, this proceedings). Does this mean we are seeing 
absorption? No. An intrinsic blazar \g-ray spectrum of this type 
is exactly what one expects (e.g., see next section) in an SC
emission model. 

To next order, the  DEBRA is  better described 
as the sum of two  emission components:  
starlight from galaxies peaking at $\sim 1$ eV, and
dust re-emission peaking at $\sim 100 \mu$ (see Fig. 1a),
The $ 1-10 \mu$ side
of the ``valley'' between the DEBRA emission peaks is typically 
a power law with  $\alpha \sim 1.$ At the 
corresponding $E_\gamma\sim 1-10$ TeV, roughly the energy range
of current TeV detectors,
$\tau_{\gamma\gamma}$ is thus $\sim$ constant! 
The shape of the spectrum is unchanged, and if we only
had the $\sim 1-10$ TeV data,  we again cannot infer absorption
(even if it is strong).
We demonstrate this explicitly in Fig. 1b by computing the
intrinsic source spectrum corrected for absorption effects. 
(The numerical calculations shown use
the exact energy dependent cross-section for $\gamma$-$\gamma$
pair production.)
Between $\sim
1-10$ TeV, both the  corrected and uncorrected spectra are
 very reasonable looking SC spectra, and without more
information (next section), we have no constraint.
Note that  
recent results at $\epsilon_\ast \sim 3 \mu$
\cite{dwekb} give a high DEBRA energy density,
$u(3\mu)\sim 2\times 10^{-3}.$ 
Even for Mrk 501
($z_s = 0.034$),
$\tau_{\gamma\gamma} \approx 0.5$ at $E_\gamma \sim 3$
TeV, i.e., absorption corrections  may in fact be important
($I_0/I \sim 2)$!

\begin{figure*}
\centerline{\psfig{file=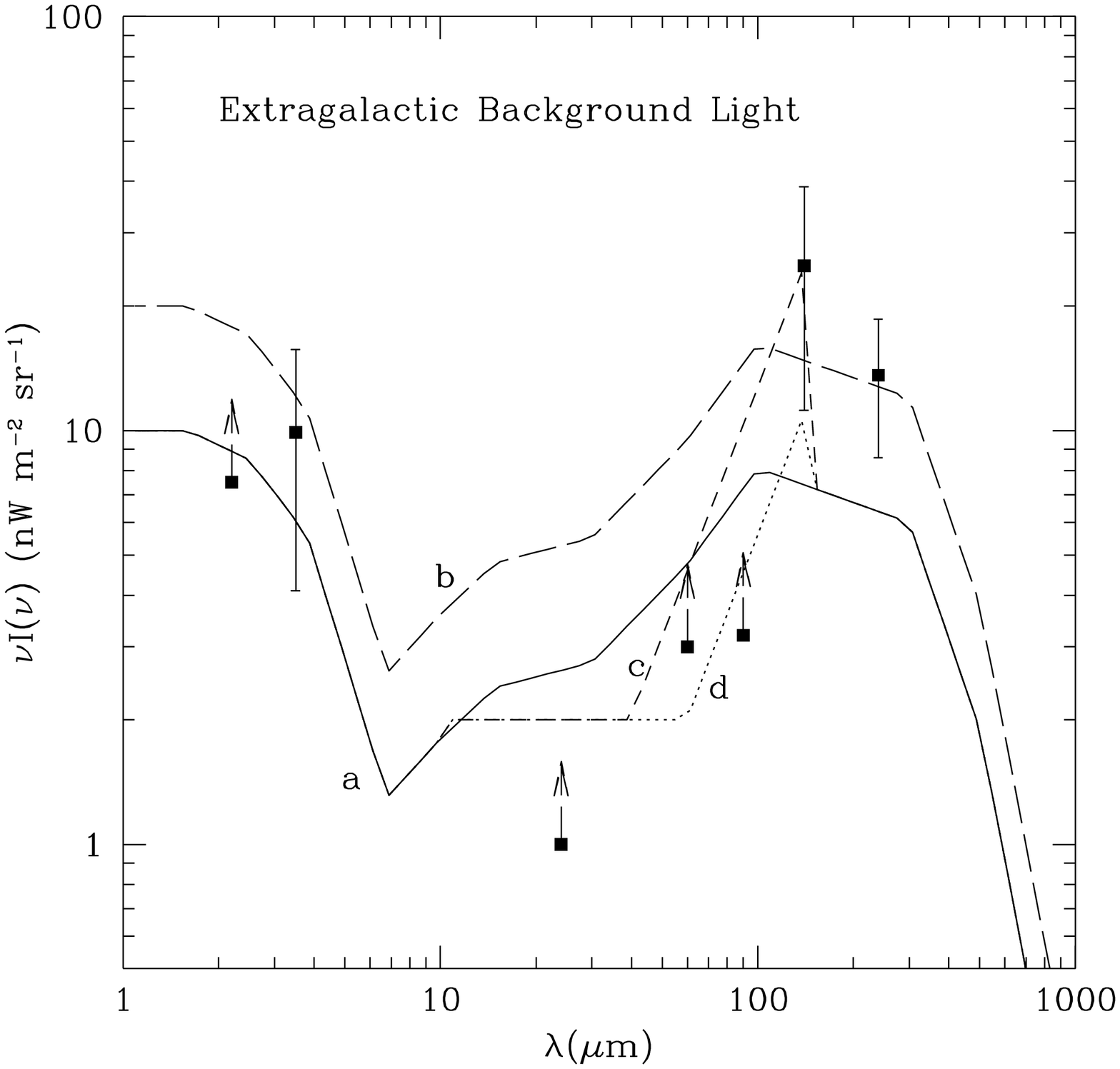,height=8cm}\psfig{file=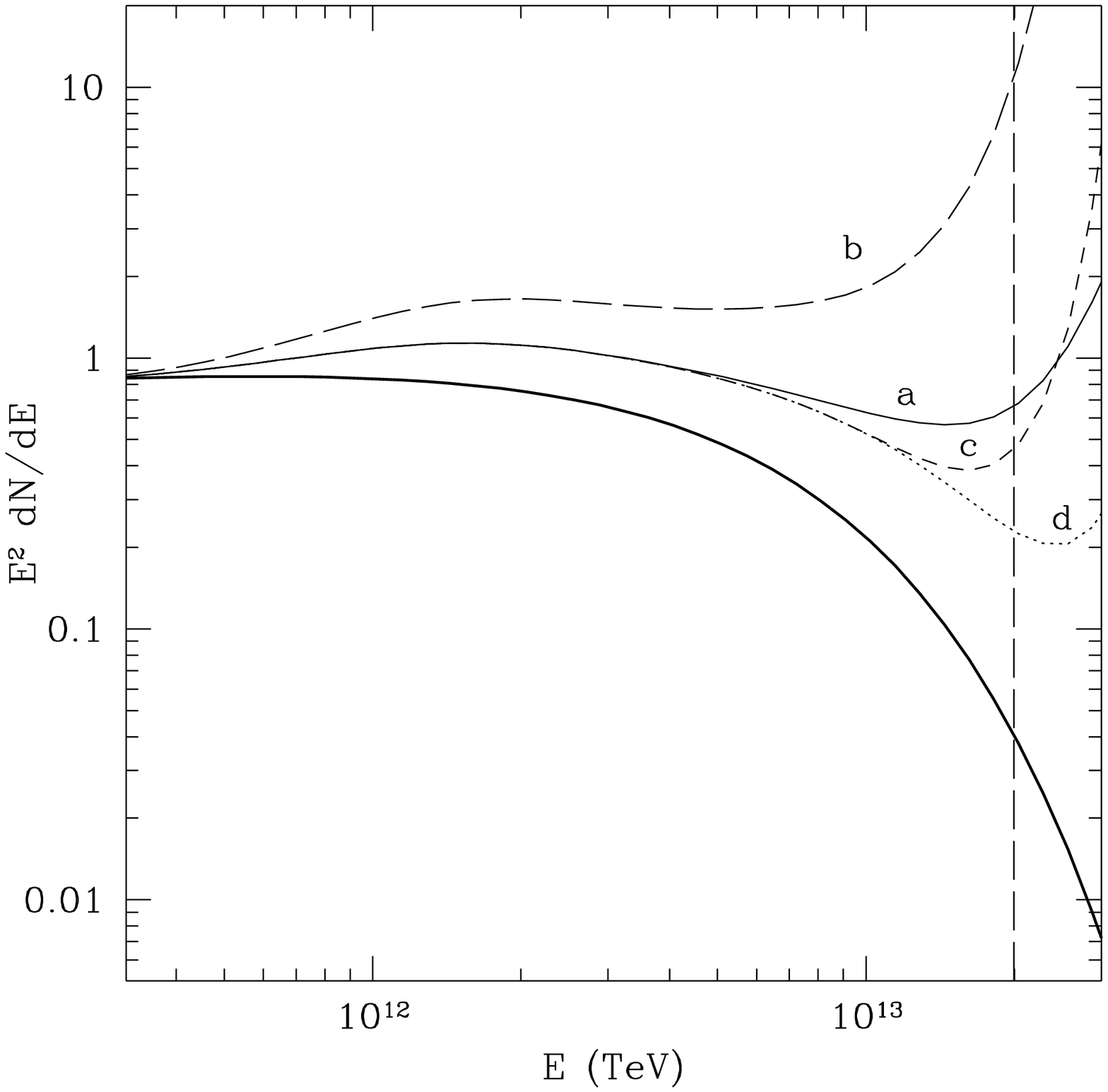,height=8cm}}
\caption{ {\it(a-left panel)} 
The DEBRA assumed for the absorption calculations 
shown in the adjoining panel.  The two rightmost data points (black squares)
are the COBE detections with 2$\sigma$ errorbars
shown. The leftmost point with errorbars (2$\sigma$) is the 
recent 3.5 $\mu$ result \cite{dwekb}.
The remaining data points are various lower limits taken from
the compilation of Dwek et al. \cite{dwek}, see  their Fig. 8.
Curve $a$ shows the DEBRA model
of Franceschini et  al. \cite{fran}. Curve $b$
shows the same model, but multiplied by
a factor 2. Curves $c$ and $d$
show two modifications 
to the DEBRA prediction in the 10-140 $\mu$ range which
are  normalized respectively to match the COBE 140 $\mu$ best 
fit value and the 2$\sigma$ lower limit. {\it (b-right panel)} The
{\it heavy (lower) solid line} shows the best fit spectrum reported by
HEGRA for
Mrk 501 (Konopelko, these proceedings). The other 
curves show the intrinsic
blazar spectra obtained by correcting this observed spectrum
for absorption  caused by the corresponding backgrounds in the left
panel. The dashed vertical line at 20 TeV represents a conservative
estimate for the maximum detected photon energy. 
}
\label{f1}
\end{figure*}

Without detailed spectral information, 
 the strongest DEBRA constraints may in fact
come from energies $E_\gamma \sim 10-30$ TeV, which
probe DEBRA energies on the ``other''
side of the valley ($\epsilon_\ast\sim10-60\mu,$). Here, 
$\alpha > 2$ and absorption should
grow  {\it super-exponentially} with \g-ray
energy.  HEGRA does {\it not} show such a rapidly falling
spectrum. The implications of this for the background models
of Fig. 1a  are shown in Fig. 1b. As a representative example
of current DEBRA models, we took that of Franceschini et al. 
\cite{fran}. This model significantly underpredicts the COBE flux
points, yet it is still marginally ruled out by the fact that 
the ``unabsorbed'' spectrum begins to curve up at $\sim$ 18 TeV,
and at 25 TeV exceeds a power law extrapolation from lower energies
by a factor 3. If the SC model applies, this is 
a serious problem since the shape of the intrinsic Compton \g-ray
spectrum is generically concave down. 
%spectrum is generically concave down. (Note that for Mrk 501,
%GRO only gives a very weak detection in the  MeV/GeV, i.e.,
%the synchrotron spectrum and thus the underlying 
%electron distribution must have a high energy cutoff. Also,
%a spectrum like the corrected one will run into
%energetics problems.)
Arbitrarily increasing the Franceschini et al. \cite{fran}
DEBRA by a factor 2, we obtain a model that fits the COBE data
well and appears compatible
with current upper limits at other energies. However, the unabsorbed spectrum
then explodes above $\sim$ 10 TeV and almost certainly is not
compatible with the intrinsic \g-ray spectrum of Mrk 501.
Most of the DEBRA models shown
in the compilation of Dwek et al. \cite{dwek} have similar problems.
To avoid them, the predicted $10-40\mu$ DEBRA flux must be
low (a few  nW ${\rm m}^{-2} {\rm sr}^{-1}$). To match the COBE points, 
the DEBRA at longer wavelengths must then increase rapidly 
with wavelength. For example (see curves $c$ and $d$ in Fig. 1), if we
assume the DEBRA spectrum shortward of 140 $\mu$ is a power law
$n(\lambda) \propto \lambda^{\alpha}$, then $\alpha$ must be $\simgreat
 4.$ Since the DEBRA is an integral (smoothed) quantity,
the individual spectra of the objects that dominate the DEBRA must
be at least as steep.
Either the standard DEBRA/galaxy evolution scenario is 
correct  and the COBE and/or HEGRA measurements
are wrong (favored by Stecker,
this proceedings), or the far IR DEBRA
is produced by objects that are not typical of what
we see in our local Universe. An increasingly discussed possibility 
(see Primack, these proceedings)
is that much of the star formation
in the Universe in fact occurs in heavily obscured regions,
e.g., in ultraluminous IR galaxies like Arp 220
which are relatively rare today. If the COBE and TeV data
are both correct, this conclusion becomes inescapable. Note that
if the IR DEBRA sources are like Arp 220 (with an IR emission
peak at $\sim 60\mu$), the 20 TeV absorption constraint
tells us they must evolve very strongly with
redshift ($\simgreat (1+z)^{3}$) and  emit the bulk of 
their light at $z\sim 3.$

\section{Predicting the Intrinsic Gamma-Ray Spectrum of a TeV Blazar}

If the SC model works for TeV blazars,  we may be able
to  robustly predict their intrinsic \g-ray spectra. The key is that the
Compton scattering responsible for the TeV \g-rays probably 
occurs in the Klein-Nishina limit. In this regime,
the target photon comes away with essentially all the energy of the incident
electron.  Also,  the cooling of the 
electrons is dominated by synchrotron radiation.  
In an external acceleration scenario
(where electrons are injected into the source region), this
means the {\it only} way to change the shape of the cooled electron 
distribution is to change the shape of the electron 
injection function, i.e., the cooled electron spectrum may
be rather insensitive to changes in the source. Since the
scattering electron is effectively replaced by a photon of 
the same energy, this also means the observed TeV 
\g-ray spectrum is essentially the cooled GeV/TeV
electron distribution (see dotted line in
Fig. 2) and, hence, may be similarly insensitive
to source changes. In 
particular, it does not depend on the target photon
distribution, as shown in Fig. 2, where a completely different
(non-synchrotron) target photon distribution gives the same Compton 
upscattered spectrum. This may explain the remarkable stability of the
Mrk 501 TeV spectrum despite the large changes seen in source 
luminosity (and may mean that coadding spectra to improve
statistics, e.g., Konopelko, this proceedings, is justified).
In short,  if we  can ``invert'' the observed synchrotron X-ray spectrum
to obtain the underlying electron distribution
(e.g., as in Fig. 2), we have  all we need to
predict the shape of the upscattered TeV spectrum. 
Extrapolating from the  spectrum observed at low energies
where intergalactic absorption should  not be
important (e.g., 500-700 GeV), we 
then  predict the unabsorbed flux at TeV energies.
Our accuracy  is limited
by  uncertainties in  $B$ and $\delta$ (the region's
characteristic magnetic field and Doppler boost factor) and the
presence of external IR target photons (too many low
energy ones take the model out of the Klein-Nishina
regime). However, bad estimates of $B$ and $\delta$ only
cause an overall energy
shift of the predicted \g-ray spectrum by a factor 
$({\delta \over B})^{1/2},$  i.e., a fairly weak dependence.
Also, the rest frame energy density of external IR photons must 
exceed  the synchrotron photon energy density to cause
significant deviations in the predicted spectrum. 
This is possible, but not likely since objects like
Mrk 421/501 have underluminous accretion disks and 
and do not have a \g-ray (Compton) luminosity that 
significantly exceeds the X-ray (synchrotron) luminosity.

\begin{figure*}
\centerline{\psfig{file=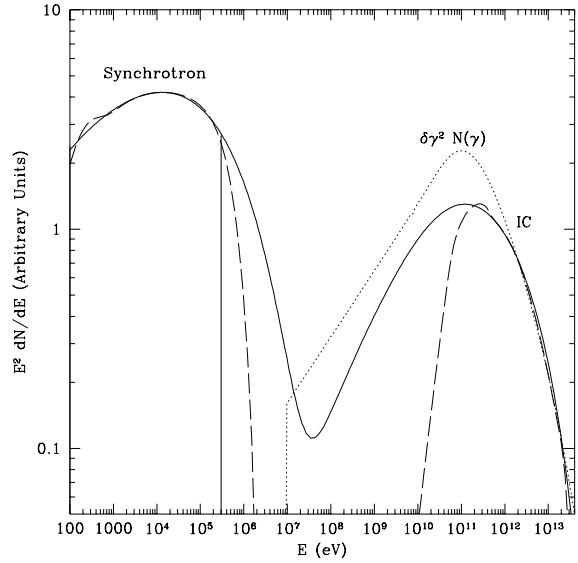,height=8cm}}
\caption{
%{\it(a-left panel)} 
The {\it solid} line is the 
time-integrated photon 
spectrum from a variable SSC model
chosen to give spectra similar to those seen in
the April 1997 Mrk 501 flare \cite{catan,pian}. (In this model, the variable
parameter is total electron luminosity; 
electrons are always injected into the source with the same
energy spectrum.)  The {\it dashed } line shows the synchrotron 
and Compton fluxes  produced by the electron distribution
reconstructed from the ``observed''
0.1-300 keV model X-ray spectrum. The target soft photon
distribution used to compute the Compton spectrum was
$n(\epsilon) \propto \epsilon^{-2}$ between $0.2<\epsilon<5$ eV
(as measured in the source frame). The {\it dotted} line shows
the electron distribution in a steady-state SSC model with
the same mean parameters as the variable SSC model.
The distribution is plotted in the same way as the photon
distribution, i.e., as $\gamma^2 N(\gamma)$ ($\gamma$
is the electron Lorentz factor.)  Above $\sim 1$ TeV,
note the excellent agreement with the Compton \g-ray spectrum.
The distribution has {\it not} been rescaled.
%{\it(b-right panel)} The relative lag/lead  of the flux at an 
%observed energy, $E_{obs},$ versus the 
%observed flux at 0.3 keV. The curves are computed 
%by running a cross-correlation analysis and plotting
%the time of the cross-corelation peak. Curve (b) shows
%the results for the variable SSC model used in the left
%panel. Curve (a) shows the results obtained when the 
%high-energy cutoff in the injected electron spectrum
%varies with electron luminosity. Curve (c) shows the results
%when the source magnetic field 
%strength instead varies with electron luminosity.
\label{f2}}
\end{figure*}
%\begin{figure*}
%  \vspace{6cm}
%\caption{A Caption}
%\label{f3}
%\end{figure*}
%\begin{figure*}
%  \vspace{6cm}
%\caption{A Caption}
%\label{f4}
%\end{figure*}

\section{Conclusions}

TeV blazars like Mrk 421/501  sources provide ideal
laboratories to test in detail the emission models for these
objects. If we can show that
a simple SC model works during at least the strongest flares,
then we can use good broadband X-ray spectra of these sources to
infer their intrinsic TeV spectra.
Then, and only then, can we look for evidence of \g-ray
absorption below $\sim 10$ TeV and attempt to constrain the 
corresponding $\sim 1-20 \mu$ DEBRA.
(Blazar modelers should also not forget that the Compton
spectra they are trying to fit could be strongly attenuated!)
Above $\sim 10$ TeV, absorption is expected to grow so 
rapidly with \g-ray energy that simply requiring  the 
absorption-corrected spectrum to be concave down is sufficient
to impose very interesting constraints on the DEBRA.
If the COBE and HEGRA Mrk 501 data are correct, the DEBRA
must rise very steeply ($n(\lambda) \propto \lambda^{\alpha}$,
with $\alpha \sim 4$)
longwards of $\sim 40 \mu.$ Unless we identify closer sources
at 20-30 TeV, the finite energy resolution of
detectors will prevent us from obtaining much stronger constraints
than those presented for this wavelength region.
In any single observation,  absorption
effects could be due both to intrinsic blazar
IR/O photons as well as intergalactic ones.
While these contributions can be difficult to disentangle
(note, though,  that internal absorption does not affect the concave
down argument),  
Mrk 421 and 501 conveniently have the
same redshift. Thus, we can require that any
absorption attributed to
intergalactic photons  be exactly the same
for {\it all} flares in {\it both} sources.
These two sources alone may  give us
the first firm handle on DEBRA \g-ray absorption.

\end{document}